\documentclass[onecolumn,12pt]{article}%
\usepackage[latin1]{inputenc}
\usepackage{graphicx}
\usepackage{amsmath}
\usepackage{cite}%
\usepackage{amsfonts}%
\usepackage{amssymb}
\newcommand \be{\begin{equation}}
\newcommand \ee{\end{equation}}

\begin{document}

\title{Electrostatics of a finite-thickness conducting cylindrical shell: coupled
elliptic-kernel integral equations}
\author{J. Ricardo de Sousa\\Universidade Federal do Amazonas,\\Departamento de F\'{\i}sica, \\3000, Japiim, 69077-000, Manaus-AM, Brazil.}
\date{}
\maketitle

\begin{abstract}
We develop an exact electrostatic formulation for a finite-length conducting
cylindrical shell of finite thickness separating two dielectric media with
arbitrary permittivity contrast. The boundary-value problem is reduced to a
coupled system of singular integral equations with elliptic kernels governing
the induced surface-charge densities on the inner and outer faces.
High-accuracy numerical solutions are combined with a systematic asymptotic
analysis that elucidates the interplay between geometry, thickness, and
dielectric contrast. All classical limiting regimes are recovered, including
the slender-body limit, the short-cylinder (ring-like) asymptote, and the
thick-shell regime dominated by the outer surface. We demonstrate that the
logarithmic short-cylinder behavior of zero-thickness models is a singular
feature, which is regularized for any finite thickness, giving rise instead to
a finite capacitance plateau. The asymptotic structure of the coupled
equations explains both the electrostatic decoupling of the inner cavity in
the thick-shell limit and the redistribution of charge between the two
surfaces. The results provide exact benchmarks for finite cylindrical
conductors, bridging classical analytical treatments and modern numerical
approaches, and furnish a high-accuracy reference solution for the validation
of axisymmetric electrostatic solvers.

\end{abstract}

\section{Introduction}

The electrostatics of finite conductors has long provided a canonical
benchmark for both analytical methods in potential theory and high-accuracy
numerical solvers. Among such geometries, the finite right-cylindrical
conductor plays a particularly prominent role. It interpolates continuously
between two singular limits: a slender-body regime, in which the surface
charge density varies weakly along the axis except near the ends, and a
short-cylinder or ring-like regime, where a vanishing length at fixed radius
produces behaviour dominated by rim effects. This mixture of axial symmetry
and sharp edges makes finite cylinders a stringent test case for any method
that seeks to resolve both global fields and localized edge singularities.

The capacitance and surface-charge distribution of finite-length cylinders
constitute a classic electrostatic problem that admits no closed-form solution
in elementary functions and has therefore been investigated extensively by
numerical and semi-analytical means. Early experimental and theoretical work
goes back to classical capacitance measurements and bounds in the tradition of
Cavendish and Maxwell\cite{c1,maxwell}. Since then, the finite-cylinder
geometry has remained a standard benchmark, combining the relative simplicity
of axial symmetry with a nontrivial edge singularity that challenges both
analysis and computation\cite{c2}.

As a result, hollow finite cylinders have attracted sustained attention in the
applied-electromagnetics and electrostatics communities. Mid-twentieth century
studies employed boundary-integral and matrix methods to generate numerical
data and practical approximations over wide ranges of aspect ratio. In this
context, the work of Vainshtein and related contributions by Kapitsa, Fock,
and collaborators\cite{c2a,c2b,c2c} are widely regarded as milestones in the
systematic numerical analysis of finite cylindrical conductors\cite{c3}.

Later developments included dual-integral-equation formulations, often reduced
to linear systems via Neumann-series constructions, method-of-moments
implementations, and semi-empirical parametrizations tailored to engineering
applications. Verolino, in particular, formulated the surface-charge-density
problem for a hollow metallic cylinder within a dual-integral-equation
framework and provided a detailed assessment of classical
approximations\cite{c4}.

In parallel, considerable effort has been devoted to the capacitance of the
open cylinder as a global observable. Classical engineering-level formulas and
numerical comparisons were reported early on, and subsequent work has refined
these approximations and tabulated high-accuracy values spanning both the tube
and ring limits. Scharstein's analysis of the ``capacitance of a tube'' is
frequently cited in this context, and modern reviews often list it alongside
earlier analytical and numerical results\cite{c5}. More recent studies have
proposed analytic and semi-analytic representations that combine explicit
singular terms with rapidly convergent expansions, such as Legendre-series
constructions, yielding closed or near-closed forms consistent with
established benchmarks\cite{c6,exp1,exp2,exp3,exp4,exp5,exp6,exp7,exp8}.

A convenient idealization adopted in many analytical treatments is that of a
zero-thickness conducting cylindrical shell held at a fixed potential. In this
limit, the problem reduces to determining a single axial surface-charge
density on the shell. The axial potential can be expressed through a
one-dimensional singular integral operator with a kernel involving complete
elliptic integrals, obtained by azimuthal integration of ring-to-ring Coulomb
interactions. When combined with quadrature schemes that explicitly embed the
universal square-root divergence at the rims, this formulation yields rapidly
convergent solutions for the surface-charge density and the capacitance over
broad ranges of aspect ratio, and has been recently clarified and extended in
a rigorous real-space integral-equation framework\cite{thin}. From a
mathematical standpoint, the main difficulty lies not in solving Laplace's
equation itself but in handling the mixed boundary character associated with
an open surface that possesses sharp rims and a non-uniform edge singularity.

In many practical situations, however, the conducting cylinder has a finite
thickness and separates two media of different permittivities. Examples
include finite-length coaxial capacitors and shielding tubes filled with oil
or gas in high-voltage installations, metallic liners surrounding dielectric
components in beamlines and detectors, and screening cylinders in precision
capacitance metrology. In such cases, the induced charge is no longer confined
to a single idealized surface but is redistributed between the inner and outer
cylindrical faces in a way that depends sensitively on geometry
(\textit{length}, \textit{radii}, \textit{thickness}) and on the dielectric
constant between the interior and exterior media. The presence of two coupled
surfaces also modifies the structure of the underlying integral equations,
turning the single-shell problem into a genuinely coupled system.

The aim of the present work is to provide a systematic integral-equation
formulation and analysis for this finite-thickness cylindrical shell, bridging
the gap between zero-thickness models and realistic cylindrical conductors
with dielectric contrast. Starting from a ring-integral representation of the
potential in cylindrical coordinates, we derive a coupled pair of
one-dimensional singular integral equations for the induced surface-charge
densities on the inner and outer faces of a conducting shell of inner radius
$a$, outer radius $b=a+t$ (where $t$ denotes the lateral (wall) thickness of
the cylinder) and length $L$, held at a uniform potential $V_{0}$ and
separating media of permittivities $\varepsilon_{\text{in}}$ and
$\varepsilon_{\text{out}}$. The kernels of these integral equations are
expressed in terms of complete elliptic integrals and exhibit integrable
square-root singularities as the observation point approaches the cylinder
rims. We then construct a Chebyshev-weighted collocation scheme that factors
out the singular edge behaviour by construction, leading to rapidly convergent
numerical solutions for the axial charge profiles and for the capacitance over
broad ranges of aspect ratio, thickness, and dielectric contrast.

Beyond its intrinsic interest as a nontrivial electrostatic boundary-value
problem, the present geometry provides a versatile benchmark for axisymmetric
electrostatic solvers and a natural finite-thickness generalization of the
zero-thickness shell studied in Ref.\cite{thin}. Our results clarify how
finite thickness and material contrast govern the redistribution of charge
between inner and outer faces, and how this, in turn, controls field
screening, edge enhancement, and effective capacitance in realistic finite
cylindrical conductors.

From a boundary-element perspective, the present approach corresponds to an
axisymmetric BEM reduction: analytic integration over the azimuth collapses
the 3D surface integrals to a 1D BIE posed on the axial generating line, with
kernels expressed through elliptic integrals. The main numerical challenge is
the combination of a \textbf{weakly singular self-interaction}
(near-coincident source/field points) and the \textit{geometric endpoint
singularities} induced by the open rims. We address both within a
\textbf{Nystr\"{o}m discretization} whose Chebyshev weighting explicitly
embeds the universal rim behavior, leading to high-order accuracy with
relatively few degrees of freedom.

The formulation developed here should be viewed as an exact integral-equation
representation of the continuum electrostatic problem, rather than as a
numerical boundary-element scheme. As such, it yields reference solutions
against which axisymmetric BEM and related panel-based methods may be
systematically validated.

The paper is organized as follows. In Section 2 we introduce the geometry,
governing equations, and the relevant dimensionless parameters. Section 3
derives the ring-integral representation and the associated elliptic kernels,
and presents the Chebyshev-weighted Nystr\"{o}m/collocation discretization
together with convergence diagnostics. In Section 4 we report numerical
results for the inner and outer charge profiles, screening properties, and the
capacitance as functions of aspect ratio, shell thickness, and dielectric
contrast, including the relevant asymptotic regimes and limiting behaviors.
Section 5 summarizes the main conclusions and outlines possible extensions.

\section{Physical Model and Methodology}

We consider a finite-thickness conducting cylindrical shell coaxial with the
$z-$axis, with inner radius $a$, outer radius $b$, and length $L$, extending
from $z=-L/2$ to $z=+L/2$. The metallic shell is maintained at a uniform
electrostatic potential $V_{0}$ on both its inner and outer cylindrical faces
(Fig. 1), and separates an inner medium of permittivity $\varepsilon
_{\text{in}}$ from an outer medium of permittivity $\varepsilon_{\text{out}}$.
In the absence of external fields the configuration is strictly axisymmetric,
so the induced surface-charge densities are independent of the azimuthal angle
$\varphi$ and may be written as functions of the axial coordinate alone:
$\sigma_{\text{in}}(z)$ on the inner surface $\rho=a$ and $\sigma_{\text{out}%
}(z)$ on the outer surface $\rho=b$, for $-L/2<z<L/2 $.

The objective of this work is to determine the induced surface-charge
densities $\sigma_{\text{in}}(z)$ and $\sigma_{\text{out}}(z)$ on a
finite-thickness conducting cylindrical shell, held at a constant
electrostatic potential and embedded between two dielectric media. Although
the finite cylinder is a classical configuration in electrostatics, the
presence of two coupled surfaces and a nontrivial edge singularity makes the
problem mathematically subtle and worth revisiting.

Our primary tool is a real-space formulation in which the potentials in the
inner and outer dielectric regions are expressed as nonlocal integral
operators acting on $\sigma_{\text{in}}(z)$ and $\sigma_{\text{out}}(z)$ .
These operators arise from azimuthal averaging of ring-to-point interactions
and are expressed in terms of complete elliptic integrals. The resulting
coupled integral equations for the inner and outer surface-charge densities
form the analytical core of the problem.

In cylindrical coordinates, the electrostatic potential at an arbitrary point
($\rho,\varphi,z$) n a medium of permittivity $\varepsilon$, generated by an
axisymmetric surface-charge density $\sigma_{R}(z^{\prime}) $ distributed on
the lateral surface $\rho^{\prime}=R$, can be written as
\begin{equation}
\Phi_{\varepsilon}(\rho,z)=\frac{1}{4\pi\varepsilon}\int\limits_{-L/2}%
^{L/2}\int\limits_{0}^{2\pi}\frac{\sigma_{R}(z^{\prime})Rdz^{\prime}%
d\varphi^{\prime}}{\sqrt{\rho^{2}+R^{2}+\left(  z-z^{\prime}\right)
^{2}-2R\rho\cos(\varphi-\varphi^{\prime})}},\label{c1}%
\end{equation}
From a boundary-element viewpoint, this is a single-layer (charge)
boundary-integral formulation for a Dirichlet conductor: enforcing the
prescribed potential on the inner and outer cylindrical faces yields a coupled
first-kind system for the surface-charge densities.

Because $\sigma_{R}(z^{\prime})$ is independent of the azimuthal angle, the
integral over $\varphi^{\prime}$, can be carried out analytically in terms of
the complete elliptic integral of the first kind $K(m)$, leading to%
\begin{equation}
\Phi_{\varepsilon}(\rho,z)=\frac{R}{\pi\varepsilon}\int\limits_{-L/2}%
^{L/2}\sigma_{R}(z^{\prime})\frac{K[m(z^{\prime})]}{\sqrt{\left(
\rho+R\right)  ^{2}+\left(  z-z^{\prime}\right)  ^{2}}}dz^{\prime},\label{c2}%
\end{equation}
with%
\begin{equation}
K[m(z^{\prime})]=\int\limits_{0}^{\pi/2}\frac{d\theta}{\sqrt{1-m(z^{\prime
})\sin^{2}(\theta)}},\label{c3}%
\end{equation}
and elliptic parameter%
\begin{equation}
m(z^{\prime})=\frac{4R\rho}{\left(  \rho+R\right)  ^{2}+\left(  z-z^{\prime
}\right)  ^{2}}.\label{c4}%
\end{equation}
In the present problem, this generic kernel will be used with $R=a$ and $R=b$
to construct coupled integral equations for the inner and outer surface-charge
densities on the finite cylindrical shell.

Our formulation follows the standard boundary-element approach for Dirichlet
electrostatics. After analytic azimuthal integration, the 3D operators
collapse to coupled 1D integrals along the generating line with kernels
expressed in terms of complete elliptic integrals. The potential is written as
a single-layer operator acting on the unknown surface-charge densities on the
inner and outer cylindrical faces, plus a constant reference. \ Enforcing the
conductor condition $\Phi=$const on each face leads to a coupled first-kind
single-layer BIE, which is then reduced to 1D by analytic azimuthal
integration (See Eqs.(\ref{c1a}) and (\ref{c6}) for the axisymmetric
single-layer reduction with kernels in closed form through complete elliptic integrals).

\subsection{Boundary-integral formulation}

Evaluating Eq. (\ref{c2}) at $\rho=a$ in the inner medium ($\varepsilon
=\varepsilon_{\text{in}}$) and separating the contributions from the inner and
outer cylindrical faces, one finds%
\begin{equation}
\Phi_{\text{in}}(a,z)=\frac{a}{\pi\varepsilon_{\text{in}}}\int\limits_{-L/2}%
^{L/2}\sigma_{\text{in}}(z^{\prime})\mathcal{G}_{aa}(z-z^{\prime})dz^{\prime
}+\frac{b}{\pi\varepsilon_{\text{in}}}\int\limits_{-L/2}^{L/2}\sigma
_{\text{out}}(z^{\prime})\mathcal{G}_{ab}(z-z^{\prime})dz^{\prime},\label{c1a}%
\end{equation}
where the kernels $\mathcal{G}_{\alpha\beta}(q)$ encode the ring--to--point
interaction between a source ring of radius $\beta$ and an observation point
at radius $\alpha$. Explicitly, one finds%
\begin{equation}
\mathcal{G}_{\alpha\beta}(q)=\frac{K\left[  \frac{4\alpha\beta}{\left(
\alpha+\beta\right)  ^{2}+q^{2}}\right]  }{\sqrt{\left(  \alpha+\beta\right)
^{2}+q^{2}}}.\label{c6}%
\end{equation}
where $K(m)$ is as defined in Eq. (\ref{c3}), here with $m_{\alpha\beta
}(q)=4\alpha\beta/\left[  \left(  \alpha+\beta\right)  ^{2}+q^{2}\right]  $.
The kernels are weakly singular (logarithmic) as $z\rightarrow$ $z^{\prime} $,
and the open rims induce universal endpoint singularities in $\sigma
_{\text{in/out}}$. Both features are handled by the Chebyshev-weighted
Nystr\"{o}m collocation described in Sec. 4.

Imposing the Dirichlet condition $\Phi(a,z)=V_{0}$ for $\left|  z\right|  <L/2
$ in Eq. (\ref{c1a}) and multiplying both sides by $\pi\varepsilon_{\text{in}%
}/a$ yields%

\begin{equation}
\int\limits_{-L/2}^{L/2}\sigma_{\text{in}}(z^{\prime})\mathcal{G}%
_{aa}(z-z^{\prime})dz^{\prime}+\frac{b}{a}\int\limits_{-L/2}^{L/2}%
\sigma_{\text{out}}(z^{\prime})\mathcal{G}_{ab}(z-z^{\prime})dz^{\prime
}=\frac{\pi\varepsilon_{\text{in}}}{a}V_{0}\text{.}\label{c5}%
\end{equation}

A completely analogous reasoning in the outer region, now evaluating Eq.
(\ref{c2}) at $\rho=b$ with $\varepsilon=\varepsilon_{\text{out}}$, gives%
\begin{equation}
\Phi_{\text{out}}(b,z)=\frac{a}{\pi\varepsilon_{\text{out}}}\int
\limits_{-L/2}^{L/2}\sigma_{\text{in}}(z^{\prime})\mathcal{G}_{ba}%
(z-z^{\prime})dz^{\prime}+\frac{b}{\pi\varepsilon_{\text{out}}}\int
\limits_{-L/2}^{L/2}\sigma_{\text{out}}(z^{\prime})\mathcal{G}_{bb}%
(z-z^{\prime})dz^{\prime}.\label{c1c}%
\end{equation}

Imposing the Dirichlet condition $\Phi(b,z)=V_{0}$ for $\left|  z\right|  <L/2
$ in Eq. (\ref{c1c}) and multiplying both sides by $\pi\varepsilon
_{\text{out}}/b$ yields%

\begin{equation}
\frac{a}{b}\int\limits_{-L/2}^{L/2}\sigma_{\text{in}}(z^{\prime}%
)\mathcal{G}_{ba}(z-z^{\prime})dz^{\prime}+\int\limits_{-L/2}^{L/2}%
\sigma_{\text{out}}(z^{\prime})\mathcal{G}_{bb}(z-z^{\prime})dz^{\prime
}=\frac{\pi\varepsilon_{\text{out}}}{b}V_{0},\label{c5a}%
\end{equation}
In the thin-shell limit $a=b$, the kernel $\mathcal{G}_{\alpha\beta}(q)$
reduces to the single-radius kernel used in the zero-thickness cylinder,
recovering%
\begin{equation}
\mathcal{G}(q)=\frac{K\left(  \frac{4a^{2}}{4a^{2}+q^{2}}\right)  }%
{\sqrt{4a^{2}+q^{2}}}\label{c6a}%
\end{equation}
so that Eqs. (\ref{c5}) and (\ref{c5a}) collapse to the single integral
equation of the thin-cylinder problem\cite{thin}. For $a\neq b$, Eqs.
(\ref{c5}) and (\ref{c5a}) thus form a coupled $2\times2$ system of singular
integral equations for the unknown inner and outer surface-charge densities
$\sigma_{\text{in}}(z^{\prime})$ and $\sigma_{\text{out}}(z^{\prime})$, with
kernels fully specified by the elliptic-integral expression in Eq. (\ref{c6})
and simple, geometry-dependent right-hand sides proportional to $V_{0}$.

\subsection{Chebyshev-weighted Nystr\"{o}m discretization}

Because of the strong but integrable edge divergence, a naive discretization
of Eq. (\ref{c5}) and (\ref{c5a}) converges slowly and may become numerically
ill-conditioned. To stabilize the computation and enforce the correct endpoint
behavior by construction, we introduce a Chebyshev-weighted parametrization.
Using the dimensionless coordinate $x=(2/L)z\in\left[  -1,1\right]  $, we
rewrite the surface-charge density as%

\begin{equation}
\left\{
\begin{array}
[c]{c}%
\sigma_{\text{in}}(z)=\frac{1}{\sqrt{1-x^{2}}}p_{\text{in}}(x)\\
\sigma_{\text{out}}(z)=\frac{1}{\sqrt{1-x^{2}}}p_{\text{out}}(x)
\end{array}
\right.  ,\label{c9}%
\end{equation}
where the weight $\left(  1-x^{2}\right)  ^{-1/2}$ captures the expected
square-root divergence at $x=\pm1$, while $p_{\text{in}}(x)$ and
$p_{\text{out}}(x)$ remains smooth on $\left[  -1,1\right]  $. When
convenient, we represent $p(x)$ by a truncated \textit{Chebyshev expansion in
polynomials} of the first kind $T_{n}(x)$\cite{tn1,tn2,tn3}
\begin{equation}
\left\{
\begin{array}
[c]{c}%
p_{\text{in}}(x)=\sum\limits_{n=0}^{N}c_{n}T_{n}(x)\\
p_{\text{out}}(x)=\sum\limits_{n=0}^{N}d_{n}T_{n}(x)
\end{array}
\right.  ,\label{c10}%
\end{equation}
where $N$ is the \textbf{spectral truncation order}. The coefficients
$\{c_{n},d_{n}\}$ are obtained numerically from a collocation system built on
$N_{c}$ Gauss--Chebyshev nodes in $\left[  -1,1\right]  $, as described in
Section 3.

The resulting integral equation is solved by a collocation method combined
with Gauss--Chebyshev quadrature, following standard spectral discretization
strategies for weakly singular kernels\cite{tn4,tn5,tn6}. This completes the
real-space integral formulation and numerical methodology. In the next section
we discuss the convergence properties of the Chebyshev scheme, compare the
resulting charge densities with the thin-shell limit, and analyze the
dependence of the capacitance on the aspect ratio and shell thickness.

\section{Numerical implementation and convergence}

In this section we detail the numerical implementation of the Chebyshev
collocation scheme and assess its convergence properties. We first discuss the
evaluation of the kernels $\mathcal{G}_{\alpha\beta}(q)$, with particular
attention to the near-singular regime $q\rightarrow0$, where accurate and
stable computation of the elliptic integrals is essential. We comment on
numerical stability across parameters of the resulting $2N\times2N$ linear
system as a function of the truncation order $N$, the aspect ratio
$\alpha=a/L$, and the thickness parameter $\delta=b/a$.

The discretization adopted here can be interpreted as an axisymmetric
boundary-element Nystr\"{o}m method on the generating line. Standard
axisymmetric BEM implementations typically approximate the unknown density
with piecewise-constant or piecewise-linear shape functions on axial elements
(``panels'') and enforce the boundary condition in a Galerkin or collocation
sense, requiring special treatment of near-singular self terms. In contrast,
we employ a global Chebyshev representation with a weight that captures the
universal rim endpoint behavior a priori; the integral operator is then
discretized directly by Gauss--Chebyshev quadrature at the corresponding nodes
(Nystr\"{o}m collocation). This ``spectral Nystr\"{o}m'' viewpoint
concentrates resolution where it is needed (near the rims) and delivers higher
accuracy per degree of freedom than low-order panel BEM, while preserving the
boundary-integral structure and the closed-form elliptic-kernel evaluation for
the matrix entries.

The performance of the method is validated by several complementary tests. In
the thin-shell limit $\delta\rightarrow1^{+}$, the numerical solutions for
$\sigma_{\text{in}}(z)$ and $\sigma_{\text{out}}(z)$ are shown to collapse
onto the single-surface profile of the zero-thickness cylinder\cite{thin}. In
the long-cylinder and short-cylinder regimes we compare the computed
capacitance with known asymptotic expressions, and we document spectral
convergence of the charge densities away from the edges. These benchmarks
provide quantitative evidence that the present discretization captures both
the universal edge behavior and the global geometry dependence with high accuracy.

With the transformation (\ref{c9}), i. e., $x=(2/L)z$, Eqs. (\ref{c5}) and
(\ref{c5a}) become the coupled integral equations over $x\in\left[
-1,1\right]  $%

\begin{equation}
\int\limits_{-1}^{1}\frac{p_{\text{in}}(x^{\prime})}{\sqrt{1-x^{\prime2}}%
}\mathcal{G}_{aa}(x-x^{\prime})dx^{\prime}+\delta\int\limits_{-1}%
^{1}\frac{p_{\text{out}}(x^{\prime})}{\sqrt{1-x^{\prime2}}}\mathcal{G}%
_{ab}(x-x^{\prime})dx^{\prime}=\overline{V}_{0},\label{c11}%
\end{equation}
and%
\begin{equation}
\frac{1}{\delta}\int\limits_{-1}^{1}\frac{p_{\text{in}}(x^{\prime})}%
{\sqrt{1-x^{\prime2}}}\mathcal{G}_{ab}(x-x^{\prime})dx^{\prime}+\int
\limits_{-1}^{1}\frac{p_{\text{out}}(x^{\prime})}{\sqrt{1-x^{\prime2}}%
}\mathcal{G}_{bb}(x-x^{\prime})dx^{\prime}=\frac{k}{\delta}\overline{V}%
_{0},\label{c11d}%
\end{equation}
where $k=\varepsilon_{\text{out}}/\varepsilon_{\text{in}}$ is the relative
permittivity and $\overline{V}_{0}=\frac{\pi\varepsilon_{\text{in}}}{a}V_{0}$
is the a dimensionless potential amplitude. For notational simplicity we keep
the symbol $V_{0}$ below and understand it as the rescaled quantity. The
dimensionless kernel $\mathcal{G}_{\alpha\beta}(q)$ follow from Eq. (\ref{c6})
after the change of variables and read%
\begin{equation}
\left\{
\begin{array}
[c]{c}%
\mathcal{G}_{aa}(x-x^{\prime})=\frac{K\left[  \frac{16\alpha^{2}}{16\alpha
^{2}+(x-x^{\prime})^{2}}\right]  }{\sqrt{16\alpha^{2}+(x-x^{\prime})^{2}}}\\
\mathcal{G}_{ab}(x-x^{\prime})=\frac{K\left[  \frac{16\alpha^{2}\delta
}{4\alpha^{2}\left(  1+\delta\right)  ^{2}+(x-x^{\prime})^{2}}\right]  }%
{\sqrt{4\alpha^{2}\left(  1+\delta\right)  ^{2}+(x-x^{\prime})^{2}}}\\
\mathcal{G}_{bb}(x-x^{\prime})=\frac{K\left[  \frac{16\alpha^{2}\delta^{2}%
}{16\alpha^{2}\delta^{2}+(x-x^{\prime})^{2}}\right]  }{\sqrt{16\alpha
^{2}\delta^{2}+(x-x^{\prime})^{2}}}%
\end{array}
\right.  .\label{c11e}%
\end{equation}
In the thin-shell limit $\delta\rightarrow1$, $\mathcal{G}_{aa},\mathcal{G}%
_{ab}$, and $\mathcal{G}_{bb}$ all reduce to the single-radius kernel
$\mathcal{G}(x-x^{\prime})$ corresponding to the zero-thickness
cylinder\cite{thin}.

We solve Eq. (\ref{c11}) and (\ref{c11d}) by a Nystr\"{o}m/collocation
discretization on $\left[  -1,1\right]  $ combined with Gauss--Chebyshev
quadrature tailored to the weight $\left(  1-x^{2}\right)  ^{-1/2}$. Specifically:

1. \textbf{Collocation grid}. We choose a set of $N_{c}$ collocation points
$\left\{  x_{j},j=1,2,..N_{c}\right\}  $, taken as the Gauss--Chebyshev nodes
\begin{equation}
x_{j}=\cos\left[  \frac{\left(  2j-1\right)  \pi}{2N_{c}}\right] \label{c11f}%
\end{equation}
for which the Chebyshev weight $\left(  1-x^{2}\right)  ^{-1/2}$ is naturally
incorporated in the associated quadrature rule. In principle, one could
decouple the number of quadrature nodes from the number of collocation points
and use a smaller Gauss grid, but in the present Nystr\"{o}m implementation we
simply use the same nodes for both purposes.

2. \textbf{Quadrature}. At each collocation point $x_{j}$ the integrals over
$x^{\prime}$ are approximated by Gauss--Chebyshev sums,%
\begin{equation}
\int\limits_{-1}^{1}\frac{p(x^{\prime})}{\sqrt{1-x^{\prime2}}}\mathcal{G}%
_{\alpha\beta}(x_{i}-x^{\prime})dx^{\prime}\approx\sum\limits_{j=1}^{N_{g}%
}w_{j}\mathcal{G}_{\alpha\beta}(x_{i}-\xi_{j})p(\xi_{j}),\label{c11g}%
\end{equation}
where \{$\xi_{j}$, $j=1,2,..N_{g}$\} are the Gauss--Chebyshev quadrature
nodes,\ $w_{j}$ are the correspondent Gauss--Chebyshev weights and $p$ denotes
either $p_{\text{in}}$ or $p_{\text{out}}$. This yields a fully discrete
representation of the coupled integral equations.

3. \textbf{Linear system}. The discretization produces a dense $2N_{c}%
\times2N_{c}$ block linear system for the unknown nodal values \{$p_{\text{in}%
}(x_{j})$, $p_{\text{out}}(x_{j})$, $j=1,2,...N_{c}$\},%
\begin{equation}
\left\{
\begin{array}
[c]{c}%
\sum\limits_{j=1}^{N_{c}}A_{ij}^{(\text{in,in})}p_{\text{in}}(x_{j}%
)+\sum\limits_{j=1}^{N_{c}}A_{ij}^{(\text{in,out})}p_{\text{out}}(x_{j}%
)=B_{i}^{\text{in}}\\
\sum\limits_{j=1}^{N_{c}}A_{ij}^{(\text{out,in})}p_{\text{in}}(x_{j}%
)+\sum\limits_{j=1}^{N_{c}}A_{ij}^{(\text{out,out})}p_{\text{out}}%
(x_{j})=B_{i}^{\text{out}}%
\end{array}
\right.  ,\label{c11a}%
\end{equation}
where the matrix entries $A_{ij}^{(\text{s,p})}$ collect the elliptic-integral
kernels evaluated at the node pairs ($x_{i},x_{j}$) together with the
corresponding Gauss--Chebyshev weights, and the vectors $B_{i}^{\text{in}}$,
$B_{i}^{\text{out}}$ encode the constant potentials on the inner and outer
faces. The resulting system is solved using standard dense linear-algebra routines.

In our computations we typically take $N_{c}$ of order $N$ (for example,
$N_{c}=N$ or $N_{c}=N+1$), which is sufficient to observe spectral convergence
of the reconstructed charge densities away from the edges.

4. \textbf{Dimensionless normalization}. For presentation purposes we report
the dimensionless surface-charge densities: $\widetilde{\sigma}_{\text{in}%
}(z)=a\sigma_{\text{in}}(z)/(\varepsilon_{\text{in}}V_{0})$ and $\widetilde
{\sigma}_{\text{out}}(z)=b\sigma_{\text{out}}(z)/(\varepsilon_{\text{out}%
}V_{0})$, which removes the trivial dependence on $V_{0}$ and $\varepsilon
_{0}$ and isolate the geometric dependence through the parameters $\alpha
=a/L$, $\delta=b/a$ and $\kappa=\varepsilon_{\text{out}}/\varepsilon
_{\text{in}}.$

Once the nodal values $\sigma_{\text{in}}(z)$ and $\sigma_{\text{out}}(z)$ are
reconstructed from $p_{\text{in}}(x)$ and $p_{\text{out}}(x)$, the inner and
outer charges are obtained from%
\begin{equation}
\left\{
\begin{array}
[c]{c}%
Q_{\text{in}}=2\pi a\int\limits_{-L/2}^{L/2}\sigma_{\text{in}}(z)dz\\
Q_{\text{out}}=2\pi b\int\limits_{-L/2}^{L/2}\sigma_{\text{out}}(z)dz
\end{array}
\right. \label{c10a}%
\end{equation}
and the total capacitance follows as%
\begin{equation}
C=\frac{\left(  Q_{\text{in}}+Q_{\text{out}}\right)  }{V_{0}}.\label{c10b}%
\end{equation}
The coupled integral equations are thus solved by a Nystr\"{o}m collocation
scheme based on Gauss--Chebyshev quadrature, following standard
spectral-discretization strategies for weakly singular
kernels\cite{tn4,tn5,tn6}. This provides a numerically stable and rapidly
convergent method for computing the surface-charge densities and the
capacitance of the finite-thickness cylindrical shell.

The Chebyshev-weighted Nystr\"{o}m formulation is designed to achieve rapid
convergence by incorporating the universal rim singularities analytically, so
that the auxiliary unknowns remain smooth on $\left[  -1,1\right]  $. In
practice, we assess discretization errors by monitoring the stability of
global observables, in particular the dimensionless capacitance $\widetilde
{C}=C/(2\pi\varepsilon_{\text{in}}a)$, as the number of Gauss--Chebyshev nodes
$N_{c}$ is increased. For the parameter ranges considered here, $\widetilde
{C}$ converges monotonically with $N_{c}$, and the residual numerical
uncertainty can be made negligible compared with the physical variations
discussed in the Results section. A representative convergence table for
$\widetilde{C}(N_{c})$ is reported at the end of Sec. 4.4, where capacitance
is analyzed in detail.

\section{Numerical Results}

In this section we present quantitative results for the finite-thickness
conducting cylinder obtained from the real-space elliptic-kernel formulation
discretized by a Chebyshev-weighted Nystr\"{o}m scheme. Unless otherwise
stated we set the dielectric contrast to $\kappa=\varepsilon_{\text{out}%
}/\varepsilon_{\text{in}}$. With the Chebyshev weight $\left(  1-x^{2}\right)
^{-1/2}$ built into the unknowns $p_{\text{in/out}}(z)$, the discrete
solutions converge exponentially in the number of collocation nodes $N_{c}$.
In practice, $N_{c}\approx80-140$ ensures $6-8$ significant digits for
capacitance $C$ and pointwise densities away from the last couple of edge
panels. The diagonal is treated by a local near-field averaging (implemented
as a diagonal correction of order $10^{-12}-10^{-10}$), which stabilizes the
Nystr\"{o}m matrix without affecting the reported digits. Limiting checks: (i)
As $\delta\rightarrow1^{+}$ the two faces merge and we recover the
zero-thickness cylinder; (ii) As $\delta>>1$, the outer face dominates and the
configuration approaches a single conducting disk of radius $b$ in the outer medium.

\subsection{Surface-charge densities: profiles and edge singularities}

Figure 2 displays the dimensionless surface-charge densities $\widetilde
{\sigma}_{\text{in}}(z)=a\sigma_{\text{in}}(z)/(\varepsilon_{\text{in}}V_{0})$
and $\widetilde{\sigma}_{\text{out}}(z)=\delta a\sigma_{\text{out}%
}(z)/(k\varepsilon_{\text{in}}V_{0})$ as functions of the normalized axial
coordinate $z/L$ for a fixed aspect ratio $\alpha=1/3$ and for two
representative values of the thickness parameter, $\delta=1.0$ and $4.0$.
\ The curves shown in Fig. 2 are obtained by solving the coupled integral
equations (\ref{c11}) and (\ref{c11d}) via Chebyshev collocation, leading to a
$2\times2$ block linear system for the internal and external surface densities.

The special case $\delta=1$ was analyzed in Ref.\cite{thin}, here we
generalize to arbitrary aspect ratios $\delta$, demonstrating the robustness
of the numerical scheme. In both cases, the charge density increases
monotonically as $z\rightarrow\pm L/2$, reflecting the universal square-root
divergence at the cylinder rims. This behavior is fully consistent with the
analytical endpoint structure encoded in Eq. (\ref{c9}) and confirms that the
numerical scheme correctly captures the non-integrable local field enhancement
associated with the sharp edges. For the thinner shell ($\delta=1$),
corresponding to the zero-thickness limit, the surface-charge density remains
nearly uniform over most of the central region, with significant variations
confined to narrow boundary layers near the ends. For a moderately long
cylinder with $\alpha=1/3$, this indicates that edge effects are largely
localized and do not dominate the interior electrostatics.

As the thickness parameter increases to $\delta=4.0$, the distribution
undergoes a qualitative change. While the functional form of the endpoint
divergence remains unchanged, the overall magnitude of the surface-charge
density increases along the entire cylinder. In the scaled coordinate $z/L$,
the two rims effectively approach each other as the shell thickens at fixed
radius, allowing the edge singularities to overlap more strongly.
Consequently, rim effects penetrate deeper into the interior, leading to a
systematically enhanced charge density even near the midplane $z=0$. This
illustrates the intrinsically nonlocal nature of electrostatics in finite
cylinders: increasing the radial thickness alters the global balance between
edge enhancement and bulk screening rather than producing a purely local modification.

The inset of Fig. 2 makes this trend explicit by showing the midplane
dimensionless densities $\widetilde{\sigma}_{\text{in/out}}(z=0)$ as function
of $\delta$. \ For $\delta=1$, the two densities coincide at $\widetilde
{\sigma}_{\text{in/out}}(z=0)\simeq0.26$ and the total surface-charge density
of the thin cylinder at the midplane therefore equals $\widetilde{\sigma
}_{\text{tot}}(z=0)=2\widetilde{\sigma}_{\text{in}}(z=0)\simeq0.52$. As
$\delta$ increases, the two midplane values separate: the inner density
decreases markedly, while $\widetilde{\sigma}_{\text{out}}(z=0)$ increases
approximately linearly with $\delta$. Importantly, this growth does not imply
that the physical outer surface-charge density diverges. Rather, it reflects
the chosen dimensionless scaling, which measures the outer density in units
set by the inner length scale $a$ (equivalently, it incorporates the geometric
lever arm $b=\delta a$).

This asymptotic decoupling follows from the structure of the coupled integral
equations (\ref{c11}) and (\ref{c11d}) in the thick-shell limit $\delta>>1$.
For fixed aspect ratio $\alpha=a/L$, the kernels involving the outer radius
contain the large parameter $\alpha\delta=b/L$. Away from the edges, $\left|
x-x^{\prime}\right|  =\mathcal{O}(1)$, the outer self-interaction kernel has
the form%
\[
\mathcal{G}_{bb}(x-x^{\prime})=\frac{K\left(  m_{bb}\right)  }{\sqrt
{16\alpha^{2}\delta^{2}+(x-x^{\prime})^{2}}}\simeq\mathcal{O}\left(
\frac{\ln\delta}{\delta}\right)  ,
\]
where the logarithm arises from $K(m)$ as $m\rightarrow1^{+}$. The
cross-coupling kernel between inner and outer faces satisfies%

\[
\mathcal{G}_{ab}(x-x^{\prime})=\frac{K\left(  m_{ab}\right)  }{\sqrt
{4\alpha^{2}\left(  1+\delta\right)  ^{2}+(x-x^{\prime})^{2}}}\simeq
\mathcal{O}\left(  \frac{1}{\delta}\right)  ,
\]
again up to at most logarithmic corrections. Consequently, the contribution of
the inner surface to the outer equation is suppressed by $1/\delta$, and the
outer density is asymptotically governed by an effective single-surface
problem. In physical units this yields a finite limiting midplane density,%
\begin{equation}
\sigma_{\text{out}}(z=0)\simeq A(\alpha,\kappa)+\mathcal{O}(\delta
^{-1}),\delta>>1\label{c2p}%
\end{equation}
where $A(\alpha,\kappa)$ depending only on $\alpha$ and on the dielectric
contrast $k$. When expressed in the dimensionless form used in Fig. 2,
however, one obtains%

\begin{equation}
\widetilde{\sigma}_{\text{out}}(z=0)=\frac{b\sigma_{\text{out}}(z=0)}%
{\varepsilon_{\text{out}}V_{0}}\simeq\left[  \frac{aA(\alpha,\kappa)}%
{\kappa\varepsilon_{\text{in}}V_{0}}\right]  \delta+\mathcal{O}(1),\label{c2d}%
\end{equation}
so the inset naturally displays an approximately linear growth with $\delta$.

By contrast, the inner equation (\ref{c11}) becomes a driven problem in which
the field generated by the outer surface is transmitted to the cavity only
through the cross-coupling scale $\mathcal{G}_{ab}\approx\mathcal{O}\left(
\frac{1}{\delta}\right)  $. As a result the inner midplane density is
progressively screened and decays algebraically,%

\begin{equation}
\sigma_{\text{in}}(z=0)\simeq\frac{B(\alpha,\kappa)}{\delta}+\mathcal{O}%
(\delta^{-2}),\label{c2e}%
\end{equation}
and therefore $\widetilde{\sigma}_{\text{in}}(z)=a\sigma(z)/(\varepsilon
_{\text{in}}V_{0})$ decreases as $1/\delta$. Physically, the thick-shell limit
corresponds to electrostatic shielding of the cavity: the induced charge
resides predominantly on the outer face, while the inner surface becomes
effectively passive. The inset data corroborate this decoupling, showing the
suppression of $\widetilde{\sigma}_{\text{in}}(z=0)$ as $1/\delta$ decay and
and the linear-in-$\delta$ growth of $\widetilde{\sigma}_{\text{out}}(z=0)$
hat arises from the explicit factor $b=\delta a$ in the chosen normalization.

For completeness, we also examined the influence of the dielectric contrast
$k=\varepsilon_{\text{out}}/\varepsilon_{\text{in}}$. Using the natural outer
normalization $\widetilde{\sigma}_{\text{out}}(0)=\delta a\sigma_{\text{out}%
}(0)/(\varepsilon_{\text{out}}V_{0})$, the case $\kappa=1$ is of course,
unchanged by the choice of $\varepsilon_{\text{in}}$ versus $\varepsilon
_{\text{out}}$ in the prefactor. For $\kappa\neq1$, the dependence on $\delta$
over the range explored here, $\delta\in\left[  1,4\right]  $, remains
qualitatively the same: the outer face continues to carry the dominant induced
charge while the inner midplane density is suppressed as the shell thickens.
Quantitatively, the main effect of changing $\kappa$ is to rescale the overall
magnitude of $\widetilde{\sigma}_{\text{out}}(0)$, consistent with the
appearance of $\kappa$ in the outer boundary condition (\ref{c11d}). Any
residual distortion of the $\delta-$dependence is comparatively weak in this
interval and reflects the finite coupling between the two surfaces through the
kernel $\mathcal{G}_{ab}$, which is not yet fully negligible for $\delta\leq4$.

In the asymptotic thick-shell regime $\delta>>1$, the inner--outer coupling is
parametrically suppressed, and the influence of the dielectric contrast
reduces to a simple rescaling of the outer solution; the midplane behavior is
then fully governed by an effective single-surface problem at radius $b$.

Despite this strong redistribution of charge between the two faces, the edge
behavior remains universal. In all cases, the reconstructed densities exhibit
the same square-root divergence at the rims\cite{c2},%

\begin{equation}
\widetilde{\sigma}_{\text{in/out}}(z)\simeq\frac{A_{\text{in/out}}}%
{\sqrt{\left(  L/2\right)  ^{2}-z^{2}}}\left[  1+\mathcal{O}(L/2-\left|
z\right|  )\right] \label{c12}%
\end{equation}
with numerically extracted local exponents $0.500\pm0.005$ over the entire
range of parameters ($\alpha,\delta,\kappa$) explored. In terms of the
weighted variables, this divergence corresponds to smooth functions
$p_{\text{in/out}}(x)$, which underlies the observed spectral accuracy.

The prefactors $A_{\text{in}}$ and $A_{\text{out}}$ quantify how the induced
charge is partitioned between the two surfaces. For $\delta\approx1$, they are
comparable, indicating strong coupling across the shell. As $\delta$
increases, $A_{\text{out}}$ grows while $A_{\text{in}}$ diminishes, consistent
with the integrated charges satisfying $\left|  Q_{\text{in}}\right|
>>\left|  Q_{\text{out}}\right|  $ for $\delta>>1$, with $Q_{\text{in}%
}/Q_{\text{out}}=\mathcal{O}(a/b)$ once the aspect ratio enters the
short-cylinder regime (large $\alpha$). Physically, field lines preferentially
terminate on the larger, more strongly fringing outer surface. For fixed
($\delta,\kappa$), increasing $\alpha$ mainly concentrates charge closer to
the rims without altering the universal exponent, and the edge amplitudes
approach well-defined limits as $\alpha\rightarrow\infty$, in agreement with
the capacitance saturation discussed in Sec. 4.3.

Beyond these physical trends, the main contribution of the present work is
methodological. The exact boundary-integral formulation (\ref{c5}) and
(\ref{c5a}) involves kernels expressed in terms of complete elliptic integrals
whose weak logarithmic singularities render naive discretizations
ill-conditioned. By extracting the endpoint behavior analytically through the
weighted representation $\sigma_{\text{in/out}}(z)=\left(  1-x^{2}\right)
^{-1/2}p_{\text{in/out}}(x)$, the remaining unknown functions become smooth on
$\left[  -1,1\right]  $ and can be determined accurately using collocation
with Gauss--Chebyshev quadrature. This yields a stable, rapidly convergent
numerical scheme that resolves edge-dominated electrostatics with the rim
singularities treated analytically and only a standard diagonal correction for
the weak kernel singularity, providing a high-accuracy reference solution for
finite cylindrical conductors held at fixed potential.

\subsection{Finite size-scaling}

To quantify the residual discretization error and obtain continuum-limit
estimates of the cylinder-center density, we perform a finite-size scaling
analysis of the Chebyshev--Nystr\"{o}m scheme. For each choice of geometric
and material parameters ($\alpha,\delta,\kappa$), we compute the midplane
values $\widetilde{\sigma}_{\text{in}}(0)$ and $\widetilde{\sigma}%
_{\text{out}}(0)$ from a sequence of increasingly refined collocation grids
with $N_{c}$ Gauss--Chebyshev nodes on $\left[  -1,1\right]  $. Throughout the
scaling study we keep fixed the quadrature rule implicit in the Nystr\"{o}m
discretization (i.e., Gauss--Chebyshev weights at the collocation nodes), so
that the only control parameter is the number of axial nodes $N_{c}$. The
center value is obtained by a local interpolation of the numerical solution
near $z=0$, which is robust because the density is smooth at the midplane and
all non-analytic behavior is confined to the endpoints $z=\pm L/2$ and already
captured by the $\left(  1-x^{2}\right)  ^{-1/2}$ weight.

In the asymptotic regime the discretization error is observed to be well
described by a leading $1/N_{c}$ correction,%

\begin{equation}
\widetilde{\sigma}(0)=\widetilde{\sigma}_{\infty}(0)+\frac{A}{N_{c}%
}+\mathcal{O}(N_{c}^{-2}),\label{c11b}%
\end{equation}
so that $\widetilde{\sigma}_{\infty}(0)$ can be extracted from a linear
regression of $\widetilde{\sigma}(0)$ versus $1/N_{c}$.

In practice, restricting the fit to the largest grids (e.g., $N_{c}\geq220$)
suppresses pre-asymptotic curvature and yields stable intercepts. We use these
intercepts as continuum-limit estimates, and we verify that they are only
weakly sensitive to the precise fitting window once the asymptotic range is
reached. We note that the apparent slope of $\widetilde{\sigma}(0)$ versus
$1/N_{c}$ may change sign depending on whether the within-panel quadrature
order is kept fixed or tied to $N_{c}$; in all cases, however, the
$1/N_{c}\rightarrow0$ intercept is stable and provides an equivalent
continuum-limit estimate.

Figure 3 illustrates the procedure for a fixed aspect ratio $\alpha=1/3$ and
two representative thickness values, $\delta=1$ and $\delta=4$. For $\delta=1$
the two cylindrical faces coincide and the physically relevant thin-shell
result is recovered by combining the inner and outer solutions; accordingly,
panel (a) reports the total midplane density in the thin-shell limit. For a
genuinely finite-thickness shell ($\delta=4$), panels (b) and (c) show that
the inner and outer midplane densities converge independently and with
comparable linear-in$-1/N_{c}$ behavior. The close-to-linear trends and the
small scatter confirm that the collocation system remains well conditioned in
this parameter range and that the midplane observable is not contaminated by
edge singularities.

Finite-size scaling of the cylinder-center surface-charge density
$\widetilde{\sigma}(0)$ at fixed aspect ratio $\alpha=1/3$, obtained from the
Chebyshev--Nystr\"{o}m discretization of the coupled integral equations.
Results are shown for two representative thicknesses, $\delta=1$ and
$\delta=4$.

More broadly, we find the same convergence pattern across the full parameter
ranges explored in this work. Varying the aspect ratio $\alpha=a/L$ primarily
changes the overall scale of $\widetilde{\sigma}(0)$ and the degree of end
enhancement, but does not alter the leading $1/N_{c}$ scaling once the
endpoint weight is incorporated. Likewise, changing the thickness parameter
$\delta=b/a$ modifies the relative importance of the self-interaction and
cross-coupling blocks in the $2\times2$ integral system, yet the Nystr\"{o}m
discretization remains stable from the thin-shell limit $\delta\rightarrow
1^{+}$ up to moderately thick shells, with convergence rates controlled by
$N_{c}$ rather than by \textit{ad hoc regularization}. Finally, the dielectric
contrast enters only through the right-hand side and the normalization of; for
fixed ($\alpha,\delta$) it produces a smooth rescaling of the outer density
and a mild quantitative redistribution between $\widetilde{\sigma}_{\text{in}%
}$ and $\widetilde{\sigma}_{\text{out}}$, without degrading numerical
conditioning or convergence. In other words, the method is robust with respect
to both geometric variation ($\alpha,\delta$) and material contrast $k$, and
the extrapolation $\widetilde{\sigma}_{\infty}(0)$ provides a controlled
benchmark for the coupled-shell problem.

These convergence tests are physically relevant because the surface density
sets the local normal field via $E_{n}(z)=\sigma(z)/\varepsilon$, and
therefore the computed profiles quantify both the peak field enhancement near
the rims and the degree to which thickening the shell redistributes charge
between the inner and outer faces. From a computational standpoint, the
combination of an exact elliptic-integral kernel with an endpoint-aware
Chebyshev parametrization yields a transferable numerical template: the same
discretization strategy extends straightforwardly to nonuniform boundary
potentials, segmented electrodes, or coupled conductor assemblies, where
controlling edge singularities and isolating continuum-limit observables is essential.

For $\delta=1$, the thin-shell limit considered in Ref.\cite{thin}, a related
convergence analysis was carried out using a panel-based implementation in
which the number of collocation nodes $N_{c}$ and the within-panel
Gauss--Chebyshev quadrature order $N_{g}$ were treated as independent
parameters (e.g., $N_{g}=16$ and $N_{c}=220$). In the present work we adopt a
pure Nystr\"{o}m discretization on Gauss--Chebyshev nodes, effectively setting
$N_{g}=N_{c}$, which simplifies the implementation and is natural for the
weighted formulation. The two procedures are fully consistent at the level of
the continuum-limit intercept $\widetilde{\sigma}_{\infty}(0)$: although the
finite$-N_{c}$ trends may appear with opposite slope when plotted against
$1/N_{c}$, this reflects only the sign and magnitude of the leading
discretization coefficient $A$ (and, in practice, the chosen plotting
convention for the $1/N_{c}$ axis), not a change in the limiting value. In all
cases, the extrapolated $N_{c}\rightarrow\infty$ estimates agree within the
expected subleading corrections, confirming that the convergence properties
reported here for $\delta>1$ smoothly connect to the validated $\delta=1$
benchmark of Ref. \cite{thin}.

\subsection{Capacitance}

An additional quantity of direct interest that follows directly from the
numerical solution for the surface-charge densities $\sigma_{\text{in}}(z)$
and $\sigma_{\text{out}}(z)$ is the self-capacitance of the finite cylindrical
conductor. For an isolated conductor held at a uniform potential $V_{0}$, the
capacitance provides a compact global measure of the electrostatic response
and is particularly useful for comparison with classical results and
asymptotic limits.

We define the geometry-dependent capacitance coefficient $C_{11}\equiv
C(\alpha)$ through the total charge on the lateral surfaces of the cylinder, namely%

\begin{equation}
C(\alpha)=\frac{2\pi a}{V_{0}}\left[  \int\limits_{-L/2}^{L/2}\sigma
_{\text{in}}(z)dz+\delta\int\limits_{-L/2}^{L/2}\sigma_{\text{out}%
}(z)dz\right]  ,\label{c13}%
\end{equation}
where $\delta=b/a$ denotes the radial thickness ratio of the cylindrical shell.

For convenience, we introduce the dimensionless capacitance%
\begin{equation}
\widetilde{C}(\alpha)=\frac{C(\alpha)}{2\pi\varepsilon_{\text{in}}%
a}.\label{c13a}%
\end{equation}

Because the capacitance involves an integral over the entire surface-charge
distribution, it is sensitive to both the bulk behavior and the edge
singularities near the rims, and therefore serves as a stringent test of the
numerical accuracy of the solution. Furthermore, its dependence on the aspect
ratio $\alpha=a/L$ and and on the thickness parameter $\delta=b/a$ enables a
direct assessment of the crossover between the long-cylinder and
short-cylinder regimes, thereby facilitating quantitative comparison with
known asymptotic results and previously reported benchmark values.

\subsubsection{Asymptotic Regimes}

Before turning to the full numerical solution of Eqs. (\ref{c11}) and
(\ref{c11d}), it is useful to summarize the asymptotic limits that organize
the electrostatics of the hollow cylindrical shell. The corresponding limiting
analysis for the zero-thickness open cylinder ($\delta=1$), including the
slender-body regime $\alpha<<1$ and the short-cylinder (ring-like) regime
$\alpha>>1$, as well as benchmark comparisons and convergence diagnostics, was
presented in Ref.\cite{thin} and will not be repeated here. Instead, we focus
on the genuinely new features introduced by a finite shell thickness
$\delta=b/a>1$, and on how the $\delta\rightarrow1^{+1}$ limit connects back
to Ref.\cite{thin}.

\paragraph{Slender-cylinder regime ($\alpha<<1$): unchanged leading mechanism}

For $\alpha=a/L<<1$,the electrostatics remains governed by a slender-body
mechanism. Away from the rims, the surface-charge density varies only slowly
along the axial direction $z$, and the electrostatic potential is dominated by
a logarithmic dependence on the large aspect ratio $L/a$. As a result, the
dimensionless capacitance exhibits the same leading growth with $1/\alpha$,
modulated by a slowly varying logarithmic correction, as in the thin-shell
problem discussed in Ref.\cite{thin}. Finite thickness mainly redistributes
charge between the inner and outer faces of the cylinder, but does not modify
the leading $\alpha\rightarrow0$ divergence In \textit{Maxwell's classical
form}\cite{maxwell}, this asymptotic behavior reads%

\begin{equation}
\widetilde{C}(\alpha)\simeq\frac{\left(  1/\alpha\right)  }{\ln\left(
2/\alpha\right)  -1}\text{, }\alpha<<1.\label{c14}%
\end{equation}

This result highlights the robustness of the slender-body regime: even when
the conductor possesses a finite radial thickness, the long-range nature of
the Coulomb interaction ensures that the electrostatics is controlled by the
axial scale $L$ rather than by microscopic details of the cross section. In
particular, the logarithmic factor $\ln\left(  2/\alpha\right)  $ originates
from the integration of the nearly uniform axial charge density over distances
spanning the entire length of the cylinder, and is therefore insensitive to
the redistribution of charge between the inner and outer surfaces. Finite
thickness enters only at subleading order, through geometry-dependent
corrections to the effective line-charge density, while the dominant
divergence of the capacitance as $\alpha\rightarrow0$ remains universal.

\paragraph{Short-cylinder regime ($\alpha>>1$): finite-thickness
regularization and plateau}

A qualitatively distinct electrostatic regime emerges in the opposite,
short-cylinder limit $\alpha\rightarrow\infty$. In the zero-thickness model
($\delta=1$), the lateral surface effectively collapses into a ring-like
geometry, and the capacitance exhibits the well-known slow logarithmic
dependence
\begin{equation}
\widetilde{C}(\alpha)\sim\frac{2\pi}{\ln(32\alpha)},\text{(}\alpha
\rightarrow\infty,\delta=1\text{)},\label{c14a}%
\end{equation}
as discussed in Ref.\cite{thin}. This asymptotic form was first derived by
Lebedev and Skal'skaya\cite{exp1} using the method of \textit{dual integral
equations}, and has since become a classical result in the electrostatic
theory of short hollow cylinders. From a physical standpoint, this behavior
reflects a \textit{dimensional crossover}: as the axial length $L$ becomes
much smaller than the radius $a$, the conductor transitions from a genuinely
three-dimensional object to an effectively one-dimensional charged ring.

This \textit{logarithmic law} is consistent with standard short-tube--to--ring
approximations widely reported in the capacitance literature. In particular,
several independent analyses predict a leading behavior of order $1/\ln
(\alpha)$ in the limit $\alpha=a/L>>1$. High-precision numerical studies of
hollow cylinders that explicitly probe very small values of $L/a$ confirm this
exceptionally slow decay of the capacitance in the short-length
regime\cite{exp1, exp2,exp3,exp4,exp5,exp6,exp7, exp8}.

The situation changes fundamentally when the cylinder possesses a finite
radial thickness. For any fixed $\delta>1$, the geometry no longer collapses
to a one-dimensional object as $L\rightarrow0$: the presence of the outer
surface at radius $b=\delta a$ introduces an additional transverse length
scale that remains finite in the short-cylinder limit. As a consequence, the
capacitance no longer decays logarithmically but instead approaches a
\textit{finite plateau},%

\begin{equation}
\widetilde{C}(\alpha,\delta,\kappa)\simeq\widetilde{C}_{\infty}(\delta
,\kappa)+\frac{c(\delta,\kappa)}{\alpha^{2}}+\mathcal{O}(\alpha^{-4})\text{,
(}\alpha\rightarrow\infty,\delta>1\text{)},\label{c14b}%
\end{equation}
with algebraic corrections controlled by the short-cylinder parameter
$L/b=1/(\alpha\delta)$. In practice, the \textit{plateau value} $\widetilde
{C}_{\infty}(\delta,\kappa)$ is extracted numerically by a linear regression
of $\widetilde{C}$ versus $1/\alpha^{2}$ over sufficiently large $\alpha$,
yielding stable intercepts and small residuals for all cases examined.

Importantly, the logarithmic behavior of the thin-shell model, Eq.
(\ref{c14a}), is recovered only in the singular limit $\delta\rightarrow1^{+}%
$,where the inner and outer surfaces merge and the coupled integral
formulation reduces to the single-surface equation analyzed in Ref.\cite{thin}%
. The ring-like short-cylinder asymptote is therefore a peculiarity of the
idealized zero-thickness model, whereas any fixed finite thickness $\delta>1$
regularizes the $\alpha\rightarrow\infty$ response and replaces the
logarithmic decay by a well-defined \textit{capacitance plateau}.

\paragraph{Thick-shell limit ($\delta>>1$): disk-controlled asymptote}

A other simplifying asymptotic regime emerges in the thick-shell limit
$\delta=b/a>>1$, with the inner radius $a$ and the length $L$ held fixed. In
this regime, the cavity region becomes progressively electrostatically
screened: the electric field generated by charges on the outer surface largely
shields the interior, and the induced charge on the inner face is strongly
suppressed. This behavior is already anticipated at the level of the coupled
integral equations, Eqs. (\ref{c11}) and (\ref{c11d}), where the
cross-coupling kernels between the inner and outer surfaces scale as inverse
powers of $b$ and therefore vanish asymptotically as $\delta\rightarrow\infty$.

As a result, the electrostatics becomes dominated by the outer cylindrical
surface, and the configuration approaches that of a single isolated conductor
of radius $b$ embedded in the outer dielectric medium. In this limit, the
detailed structure of the cavity is immaterial, and the capacitance is
controlled entirely by the outer geometry. The corresponding short-cylinder
plateau therefore tends toward a disk-controlled asymptote.

This limiting behavior can be made explicit by invoking \textit{Kirchhoff's
classical result} for the capacitance of a thin conducting circular disk. In
his seminal analysis of electrostatic boundary-value problems, Kirchhoff
showed that an isolated conducting disk of radius $b$ in a homogeneous medium
of permittivity $\varepsilon_{\text{out}}$ has capacitance%
\begin{equation}
C_{\text{disk}}=8\varepsilon_{\text{out}}b,\label{c14d}%
\end{equation}
a result first derived in the context of potential theory in the 19th
century\cite{kirchhoff}.

Expressed in the dimensionless normalization adopted here, this immediately
yields the thick-shell asymptote%
\begin{equation}
\widetilde{C}_{\infty}(\delta,\kappa)\sim\left(  \frac{4\kappa}{\pi}\right)
\delta\text{, }\delta\rightarrow\infty,\label{c14c}%
\end{equation}
with $\kappa=\varepsilon_{\text{out}}/\varepsilon_{\text{in}}$ denotes the
dielectric contrast. Finite$-\delta$ corrections to this linear growth arise
from the residual influence of the inner surface and from axial finite-size
effects, and are governed by the small geometric ratios $a/b=1/\delta$ and
$L/b$.

\subsubsection{Capacitance Behavior Across Aspect Ratios}

Figure 4 compiles the dimensionless capacitance $\widetilde{C}(\alpha
,\delta,\kappa=1)$ over a broad range of aspect ratios $\alpha=a/L$ for
several representative thicknesses $\delta>1$. The family of curves provides a
compact global diagnostic of the coupled-surface solution and organizes
cleanly according to the asymptotic regimes summarized above (and, in the
singular thin-shell case $\delta=1$, in Ref.\cite{thin}). In the
slender-cylinder regime $\alpha<<1$, all curves rise rapidly and remain
comparatively close to one another, reflecting the fact that the leading
mechanism is still controlled by the axial scale $L$ and the associated
Coulomb logarithm; a finite wall thickness predominantly redistributes charge
between the two faces without altering the universal leading growth.

At intermediate aspect ratios, the curves become clearly ordered by $\delta$:
thicker shells systematically exhibit larger $\widetilde{C}$ at fixed $\alpha
$. This is the first unambiguous signature of genuinely finite-thickness
physics. Once $\alpha$ is no longer parametrically small, the outer surface at
$b=\delta a$ contributes more effectively to charge storage, and the cavity
becomes progressively less influential in setting the equipotential condition,
increasing the total charge required for a given $V_{0}$.

The most striking departure from the zero-thickness problem appears in the
short-cylinder regime $\alpha>>1$. For every fixed $\delta>1$, $\widetilde
{C}(\alpha)$ approaches a finite plateau $\widetilde{C}_{\infty}(\delta
,\kappa)$ as $\alpha\rightarrow\infty$, with the approach to saturation well
captured by algebraic corrections in $1/\alpha^{2}$, consistent with an
expansion in the short-cylinder parameter \ $L/b=1/(\alpha\delta)$. This
plateau is the central qualitative difference relative to $\delta=1$: the
ideal thin shell is singular in the $\alpha\rightarrow\infty$ limit and
exhibits the familiar ring-like logarithmic behavior discussed in
Ref.\cite{thin}, whereas any finite thickness retains a transverse length
scale and prevents the collapse to an effectively one-dimensional object.

The inset of Fig. 4 shows the extracted saturation value $\widetilde
{C}_{\infty}(\delta,\kappa=1)$ as a function of $\delta$ over the
thin-to-moderate thickness range explored. The monotone increase corroborates
the progressive dominance of the outer surface as the cavity becomes screened.
The upward curvature is consistent with the onset of the thick-shell tendency
discussed above: as $\delta$ increases, the inner face becomes
electrostatically irrelevant and the plateau bends toward the disk-controlled
scaling implied by the Kirchhoff limit.

\subsection{Convergence check for the capacitance}

To further substantiate the numerical reliability of the capacitance plateau
discussed above, we quantify the convergence of the dimensionless capacitance
$\widetilde{C}(N_{c})$ by a finite-size scaling analysis. As in the scaling
used previously for the center densities, Eq. (\ref{c11b}), we assume the
large$-N_{c}$ asymptotic form%

\begin{equation}
\widetilde{C}(N_{c})=\widetilde{C}_{\infty}+\frac{B}{N_{c}}+\mathcal{O}%
(N_{c}^{-2}),\label{c14h}%
\end{equation}
so that $\widetilde{C}_{\infty}$ can be estimated from a linear regression of
$\widetilde{C}(N_{c})$ versus $1/N_{c}$.

Figure 5 shows that the computed values fall on an essentially straight line
over the range $N_{c}\geq200$, for the representative parameter set
($\alpha,\delta,\kappa$)$=$($1/3,4,1$), providing direct evidence that the
leading discretization error is well captured by the $1/N_{c}$ correction. The
extrapolation yields $\widetilde{C}_{\infty}=6.870944116657182$ which we take
as the continuum estimate for this representative parameter set. This
monotonic decay, together with the linear behavior in Fig. 5, confirms that
the capacitance values reported in the remainder of this section are not
affected by residual discretization effects at the level relevant for the
physical trends discussed.

Table 1 documents the convergence of the dimensionless capacitance
$\widetilde{C}(N_{c})$ with the number of Gauss--Chebyshev collocation nodes
per surface ($N_{c}$). The relative deviation $\epsilon_{c}(N_{c})=\left|
\widetilde{C}(N_{c})-\widetilde{C}_{\infty}\right|  /\widetilde{C}_{\infty}$ s
overall decreasing, with a mild non-monotonic bump at $N_{c}=300$
($2.080\times10^{-3}$) and subsequent decay to $1.044\times10^{-3}$ at
$N_{c}=400$, $8.348\times10^{-4}$ at $N_{c}=500$, and $6.953\times10^{-4}$ at
$N_{c}=600$. This behavior is consistent with the Chebyshev-weighted
formulation: the endpoint singularity at the rims is embedded by construction
and the remaining smooth components are resolved at high order. In all
subsequent computations we choose $N_{c}$ such that the discretization
uncertainty in $\widetilde{C}$ lies well below the physical variations
discussed in the following sections.

\textbf{Table 1}: Convergence of the dimensionless capacitance $\widetilde
{C}(N_{c})$ with the number of Gauss--Chebyshev collocation nodes $N_{c}$ (per
surface) for the representative parameter case ($\alpha,\delta,\kappa$%
)$=$($1/3,4,1$) of Fig. 5. The continuum estimate is $\widetilde{C}_{\infty
}=6.870944116657182$, obtained from the finite-size scaling fit $\widetilde
{C}(N_{c})=\widetilde{C}_{\infty}+B/N_{c}$. The relative deviation is
$\epsilon_{c}(N_{c})=\left|  \widetilde{C}(N_{c})-\widetilde{C}_{\infty
}\right|  /\widetilde{C}_{\infty}$. Values show an overall $\mathcal{O}%
(1/N_{c}){}$ trend with a small non-monotonic bump at $N_{c}=300$.

\begin{center}%
\begin{tabular}
[c]{lll}\hline\hline
$N_{c}$ & $\widetilde{C}(N_{c})$ & $\epsilon_{c}(N_{c})\times10^{-3}%
$\\\hline\hline
$200$ & $6.88523245917$ & $1.390$\\
$300$ & $6.88049614458$ & $2.080$\\
$400$ & $6.87811478665$ & $1.044$\\
$500$ & $6.87668018018$ & $0.8348$\\
$600$ & $6.87572129792$ & $0.6953$\\\hline\hline
\end{tabular}
\end{center}

In summary, the capacitance results in this section provide a coherent picture
across geometry and materials. After embedding the rim singularity, the
finite-size scaling in Eq. (\ref{c14h}) produces strictly monotone sequences
$\widetilde{C}(N_{c})$ and robust continuum estimates $\widetilde{C}_{\infty}%
$; already at $N_{c}=400-500$ the discretization error falls well below
$10^{-3}$ (cf. Table 1). In the \textbf{slender regime} ($\alpha>>1$)
thickness plays a weak role and $\widetilde{C}$ follows the classical
logarithmic trend of the zero-thickness model. In contrast, for \textbf{short
cylinders} ($\alpha<<1$) any \textbf{finite thickness} ($\delta>1$)
regularizes the response into a \textit{finite plateau} governed by the outer
radius $b$. The plateau increases monotonically with $\delta$ and tends to the
\textit{Kirchhoff disk limit} as the outer face dominates. Changing the
dielectric contrast $\kappa=\varepsilon_{\text{out}}/\varepsilon_{\text{in}}$
primarily rescales magnitudes through while preserving these qualitative trends.

The results in Table 2 summarize our continuum Nystr\"{o}m--Chebyshev
estimates of the dimensionless capacitance $\widetilde{C}(N_{c})$ across
slender, intermediate, and short-cylinder regimes. Values are obtained from a
two-point finite-size fit using $N_{c}{}=300,400$. These data provide compact
benchmarks and will serve as the reference set for the parametric studies that follow.

\textbf{Table 2}: Dimensionless capacitance $\widetilde{C}(N_{c})$ (same
normalization as in Figs. 4--5). All values are Nystr\"{o}m--Chebyshev
continuum estimates obtained from a two-point finite-size fit with $N_{c}%
=300$, $400$. The last two columns report the raw values used in the fit.

\begin{center}%
\begin{tabular}
[c]{lllllll}\hline\hline
Case & $\alpha=L/a$ & $\delta=b/a$ & $\kappa=\varepsilon_{\text{out}%
}/\varepsilon_{\text{in}}$ & $\widetilde{C}(N_{c}\rightarrow\infty)$ &
$\widetilde{C}(N_{c}=300)$ & $\widetilde{C}(N_{c}=400)$\\\hline\hline
A & $6.00$ & $1.05$ & $1.0$ & $1.30733541009$ & $1.30815492632$ &
$1.30795004727$\\
B & $1.00$ & $1.30$ & $1.0$ & $2.27300133001$ & $2.27547689078$ &
$2.27485800059$\\
C & $0.25$ & $1.50$ & $2.0$ & $7.31426851348$ & $7.35108007254$ &
$7.34187718278$\\
D & $0.25$ & $4.00$ & $2.0$ & $14.12877818098$ & $14.15433989536$ &
$14.14794946676$\\\hline\hline
\end{tabular}
\end{center}

With the $k=2$ cases (C--D), the dielectric-contrast effect is explicit: at
fixed $\alpha=0.25$, raising $\kappa$ from $1$ to $2$ nearly doubles the
continuum capacitance ($\sim1.9\times$), consistent with outer-region control
in the short-cylinder regime. For $\kappa=2$ and $\alpha=0.25$, increasing
thickness from $\delta=1.5$ to $4.0$ yields another $\sim1.9\times$ gain,
reinforcing the disk-controlled plateau picture. By contrast, the
slender/intermediate $\kappa=1$ cases (A--B) vary modestly, indicating that
dielectric contrast mainly rescales magnitudes while geometry ($\alpha,\delta
$) sets the regime. In short-cylinder geometries the outer-face--dominated
plateau of $\widetilde{C}$ and the near-linear scaling with $\kappa
=\varepsilon_{\text{out}}/\varepsilon_{\text{in}}$ provide robust knobs for
engineering capacitance targets under tight axial footprints. A brief
discussion of prospective applications is deferred to the conclusion.

\section{Conclusion}

In this work we developed a controlled numerical treatment of the
electrostatics of a finite, open, hollow conducting cylinder of finite radial
thickness $\delta=b/a$ embedded in a two-medium dielectric environment. By
recasting the boundary-value problem into coupled Chebyshev-weighted integral
equations for the inner and outer faces, we obtained a discretization that
resolves both the global nonlocal coupling and the localized rim enhancement
in a single unified scheme.

The computed surface-charge densities exhibit the universal square-root edge
divergence at $z\rightarrow\pm L/2$ across the full parameter range studied,
while the corresponding smoothness of the Chebyshev-weighted unknowns
underlies the observed spectral accuracy. Finite thickness primarily affects
the partition of induced charge between the two surfaces: for $\delta\approx1$
the inner and outer densities remain strongly coupled, whereas increasing
$\delta$ progressively shifts charge to the outer face and suppresses the
inner response, consistent with electrostatic shielding of the cavity. The
midplane data make this separation explicit and support the thick-shell
decoupling picture, in which the outer solution approaches an effective
single-surface problem while the inner density decays algebraically with
$\delta$.

Despite this strong redistribution of local charge, the capacitance provides a
compact global characterization that cleanly bridges the relevant regimes. In
the slender-body limit $\alpha<<1$ the leading divergence remains controlled
by the axial scale and is only weakly affected by thickness, while at
intermediate $\alpha$ thicker shells yield systematically larger capacitances
at fixed geometry. Most importantly, any finite thickness $\delta>1$
regularizes the short-cylinder response: as $\alpha\rightarrow\infty$ the
capacitance approaches a finite plateau $\widetilde{C}_{\infty}(\delta,k)$
with algebraic corrections, replacing the singular ring-like logarithmic
behavior that characterizes the ideal thin shell $\delta=1$ (Ref.\cite{thin}).
The extracted plateau increases monotonically with $\delta$ and shows the
expected trend toward the disk-controlled scaling at larger thickness,
consistent with the onset of the outer-surface-dominated limit.

Finally, we examined the role of dielectric contrast $k=\varepsilon
_{\text{out}}/\varepsilon_{\text{in}}$. Over the range explored, the
qualitative trends persist: thickness drives the cavity shielding and
outer-face dominance, while changing $k$ primarily rescales the outer density
through the boundary condition, with only weak residual distortions at
moderate $\delta$. These results establish the coupled-kernel
Chebyshev--Nystr\"{o}m framework as a robust benchmark tool for
finite-thickness cylindrical shells, and they provide accurate reference data
for both local field enhancement (via $\widetilde{\sigma}(z)$) and global
response (via $\widetilde{C}$) across broad geometric and material parameter ranges.

The framework developed in this work can be extended in several directions
without conceptual modification, including nonuniform boundary potentials
(segmented or biased electrodes), coupled multi-conductor configurations, and
geometries relevant to guarded-capacitance metrology and high-voltage
electrode design, where finite-thickness effects and edge-field enhancement
compete. In this broader context, the present study provides not only
quantitative benchmark data for finite cylindrical shells, but also a reusable
boundary-element computational template for end-dominated electrostatics in
axisymmetric conductor problems. In particular, the coupled elliptic-kernel
formulation and its spectrally accurate Nystr\"{o}m discretization furnish a
high-accuracy reference solution against which axisymmetric BEM
implementations based on panel discretizations can be systematically validated.

Beyond the present validation, the ($\alpha,\delta,k$) maps and the fast
axisymmetric Nystr\"{o}m--BEM solver enable practical uses in: sizing
finite-length coaxial capacitors, RF/high-voltage feedthroughs, and
UHV/cryogenic shields where the short-cylinder plateau affords tolerance to
geometric variations; compact capacitive cells/sensors exploiting outer-face
control to stabilize $\widetilde{C}$; field-shaping and guard-ring design in
gaseous/semiconductor detectors (mitigating small-pixel effects); and inverse
design of multi-shell capacitance matrices for code verification and
metrology, with the Kirchhoff-disk limit serving as a calibration reference
for edge corrections.

\section*{Declaration of competing interest}

The author declares that there are no known competing financial interests or
personal relationships that could have appeared to influence the work reported
in this paper.

\section*{Acknowledgements}

The author acknowledges the use of an AI-based language model (ChatGPT,
OpenAI) to assist in improving the English language and clarity of some parts
of this manuscript. All scientific content, results, and interpretations are
the sole responsibility of the author.

\clearpage

\begin{figure}[p]
\centering
\includegraphics[width=0.7\linewidth]{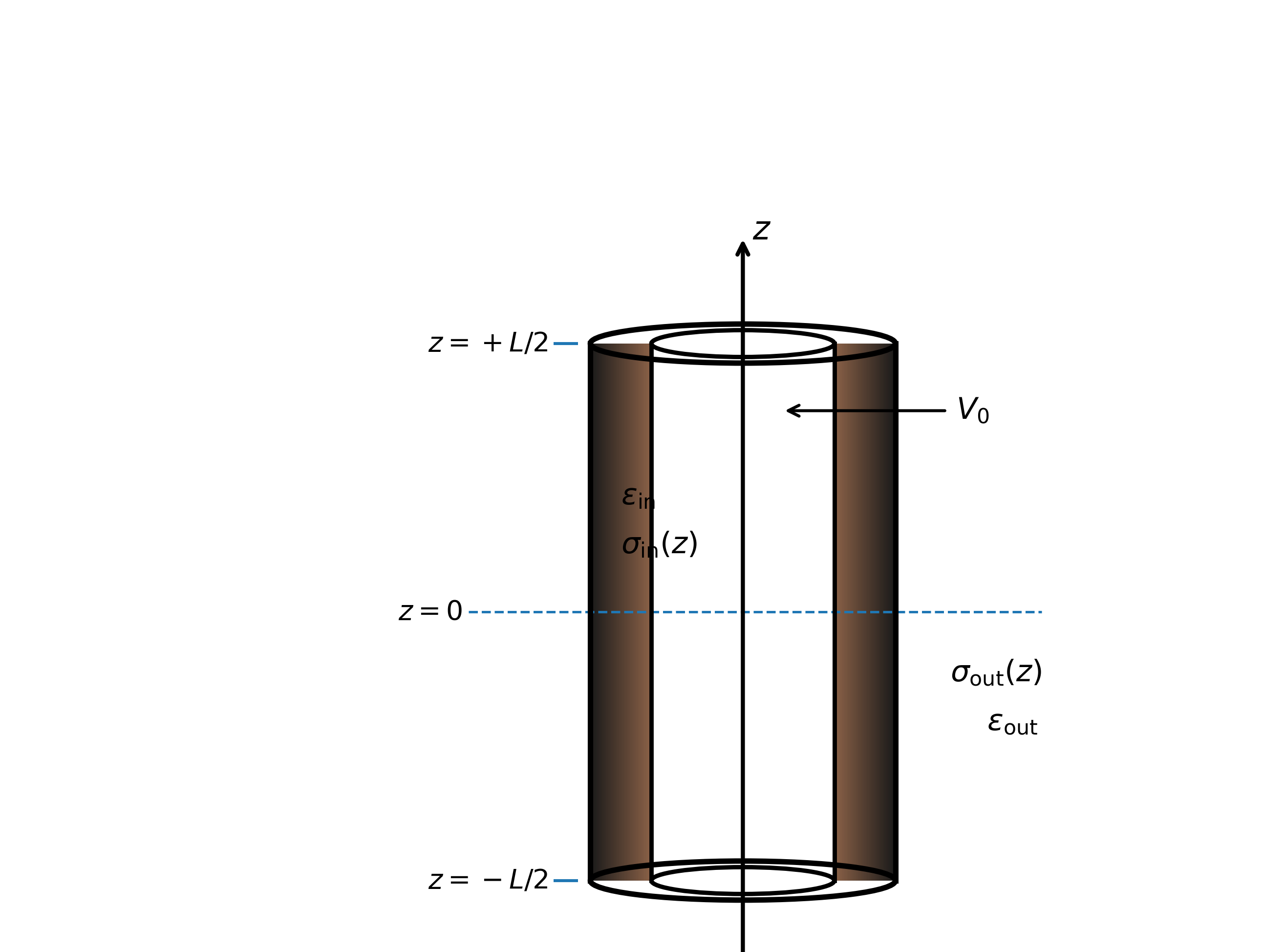}%

\caption{Schematic geometry of the finite conducting cylindrical shell. The shell has inner radius $a$, outer radius $b$, and length $L$, extending from $z=-L/2$ to $z=+L/2$. Both the inner and outer cylindrical faces are held at the same fixed potential $V_0$, separating an inner medium of permittivity $\varepsilon_{\mathrm{in}%
}$ from an outer medium of permittivity $\varepsilon_{\mathrm{out}%
}%
$. The induced surface-charge densities on the inner and outer faces are denoted by $\sigma_{\mathrm{in}%
}(z)$ and $\sigma_{\mathrm{out}}%
(z)$, respectively. Cylindrical coordinates $(\rho,z)$ are measured with respect to the symmetry axis.}%

\end{figure}

\begin{figure}[p]
\centering
\includegraphics[width=0.75\linewidth]{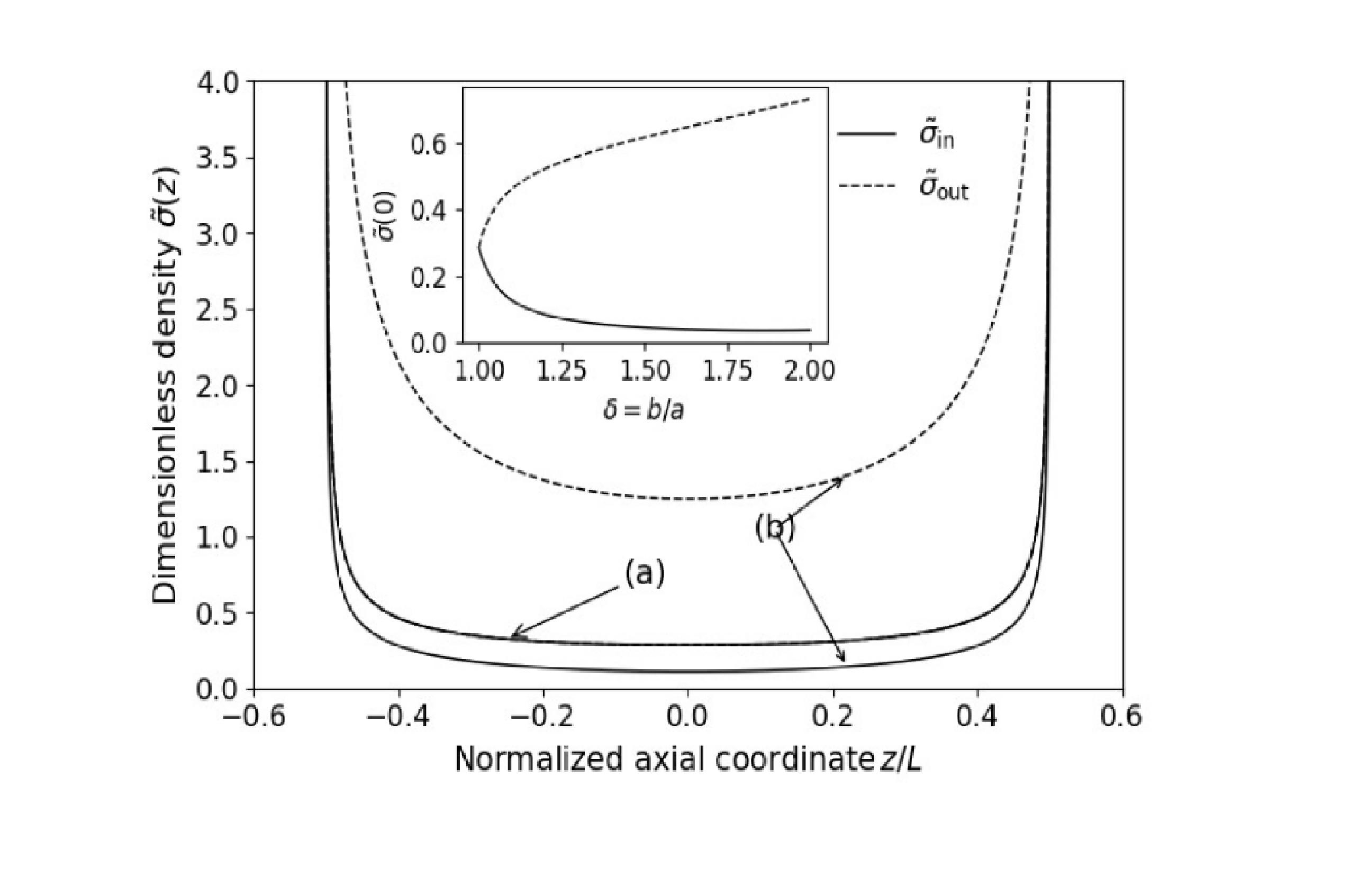}%

\caption{Dimensionless surface-charge densities $\tilde\sigma_{\mathrm{in}%
}(z)=a\,\sigma_{\mathrm{in}}(z)/(\varepsilon_{\mathrm{in}}%
V_0)$ (solid) and $\tilde\sigma_{\mathrm{out}}(z)=b\,\sigma_{\mathrm{out}%
}(z)/(\varepsilon_{\mathrm{out}}%
V_0)$ (dashed) as functions of $z/L$, for $\alpha=1/3$, $\kappa=1$, and (a) $\delta=1$ and (b) $\delta=4$. The monotonic increase toward $z=\pm L/2$ reflects the universal square-root edge divergence. The inset shows the midplane values $\tilde\sigma(0)$ versus $\delta$: the outer density saturates while the inner density decays algebraically, indicating electrostatic decoupling in the thick-shell limit.}%

\end{figure}

\begin{figure}[p]
\centering
\includegraphics[width=0.95\linewidth]{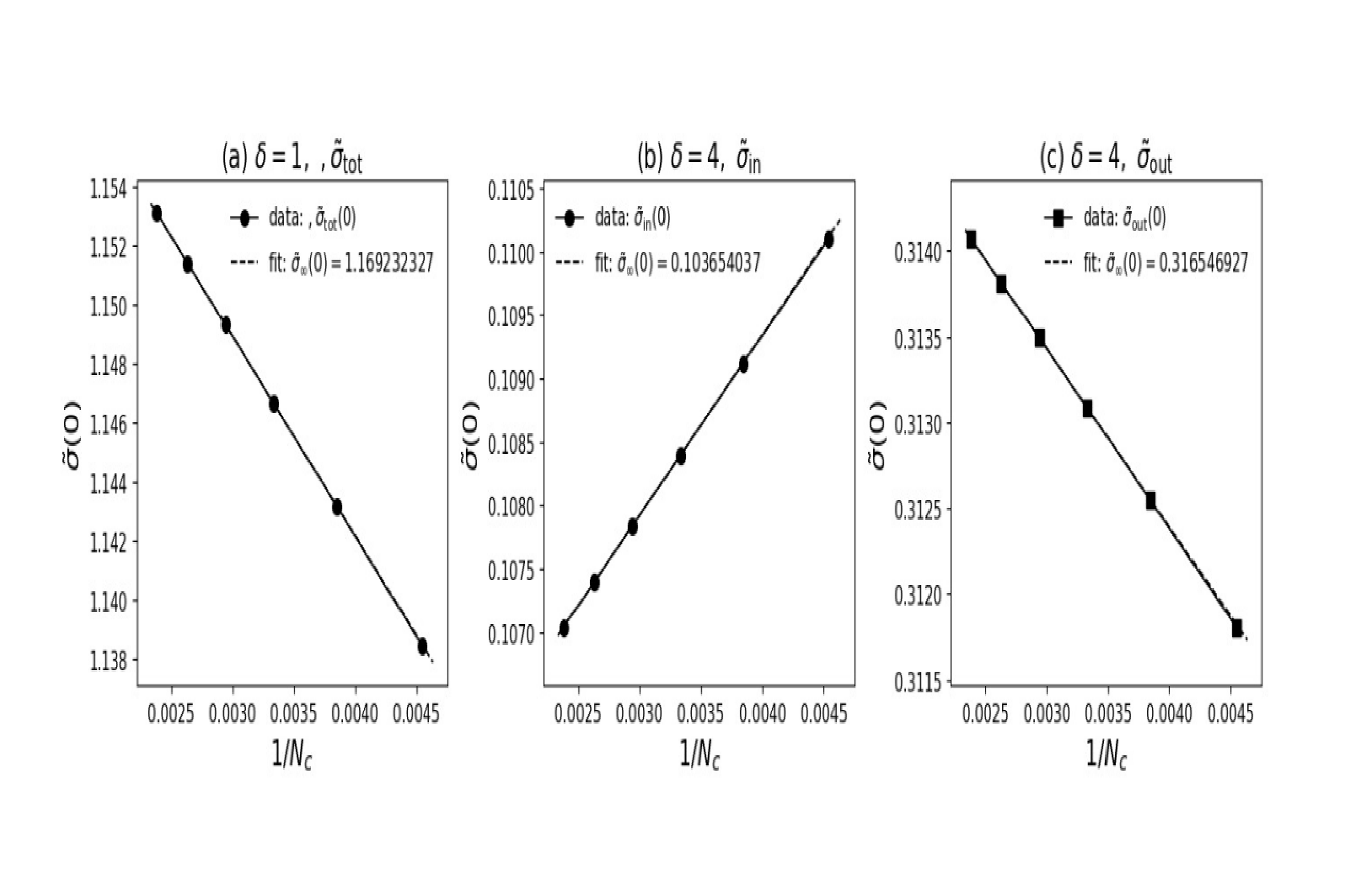}%

\caption{Finite-size scaling of the cylinder-center surface-charge density $\tilde\sigma(0)$ at fixed aspect ratio $\alpha=1/3$, obtained from the Chebyshev--Nystr\"om discretization of the coupled integral equations. Results are shown for two representative thicknesses, $\delta=1$ and $\delta=4$.}%

\end{figure}

\begin{figure}[p]
\centering
\includegraphics[width=0.75\linewidth]{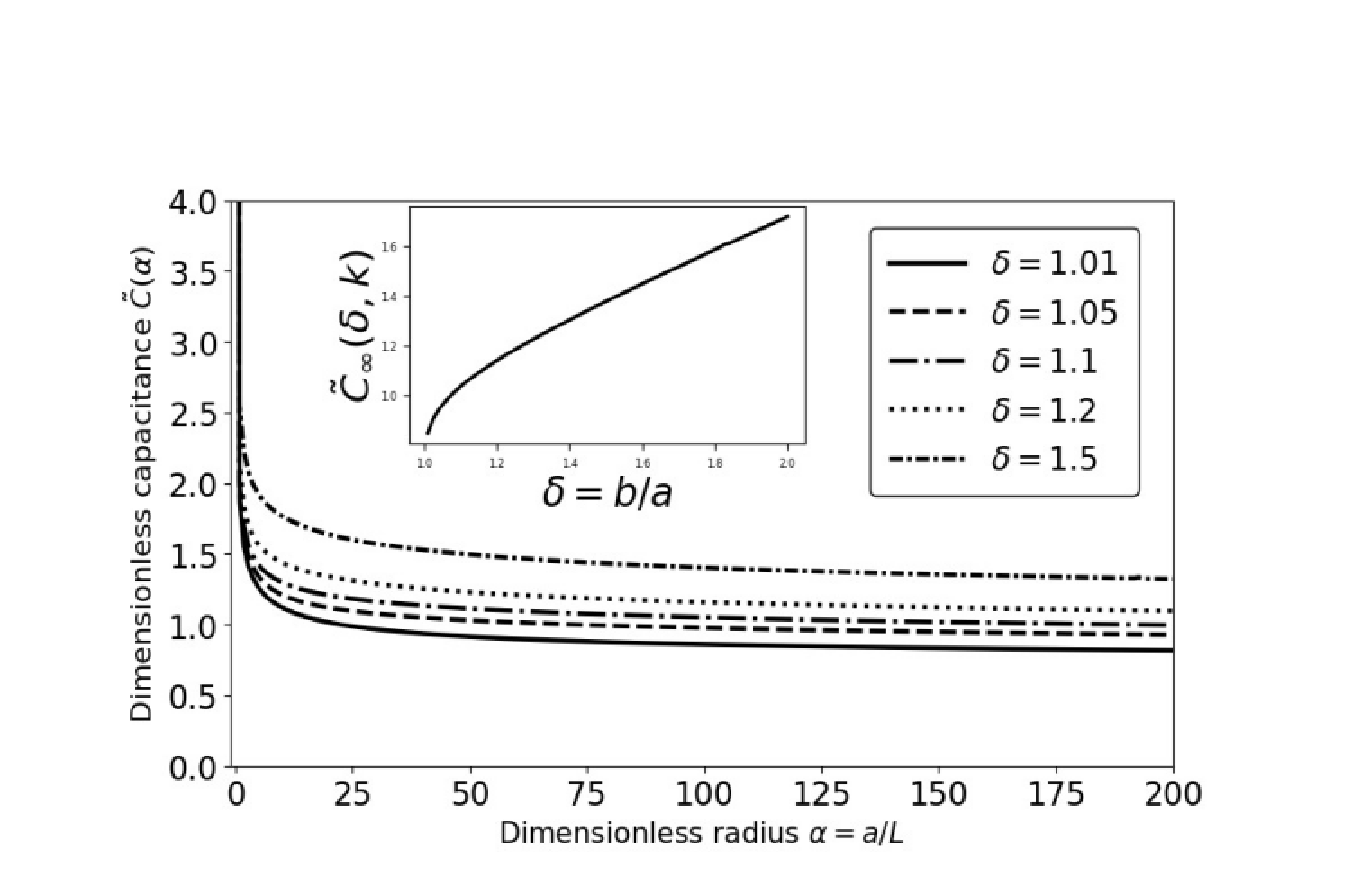}%

\caption{Dimensionless capacitance $\tilde C(\alpha)=C(\alpha)/(2\pi\varepsilon_{\mathrm{in}%
}%
a)$ of a finite cylindrical conductor as a function of the aspect ratio $\alpha=a/L$, for several values of the thickness parameter $\delta=b/a$ at fixed dielectric contrast $\kappa=\varepsilon_{\mathrm{out}%
}/\varepsilon_{\mathrm{in}}%
=1$. For $\alpha\ll1$, all curves collapse onto the slender-body regime governed by universal logarithmic growth. In the opposite short-cylinder limit $\alpha\gg1$, the capacitance saturates to a $\delta$-dependent plateau, reflecting finite-thickness regularization of the ring-like singularity present in the thin-shell model. The inset shows the asymptotic plateau value $\tilde C_{\infty}%
(\delta,\kappa)$ as a function of $\delta$, highlighting its monotonic increase toward the disk-controlled limit.}%

\end{figure}

\begin{figure}[p]
\centering
\includegraphics[width=0.65\linewidth]{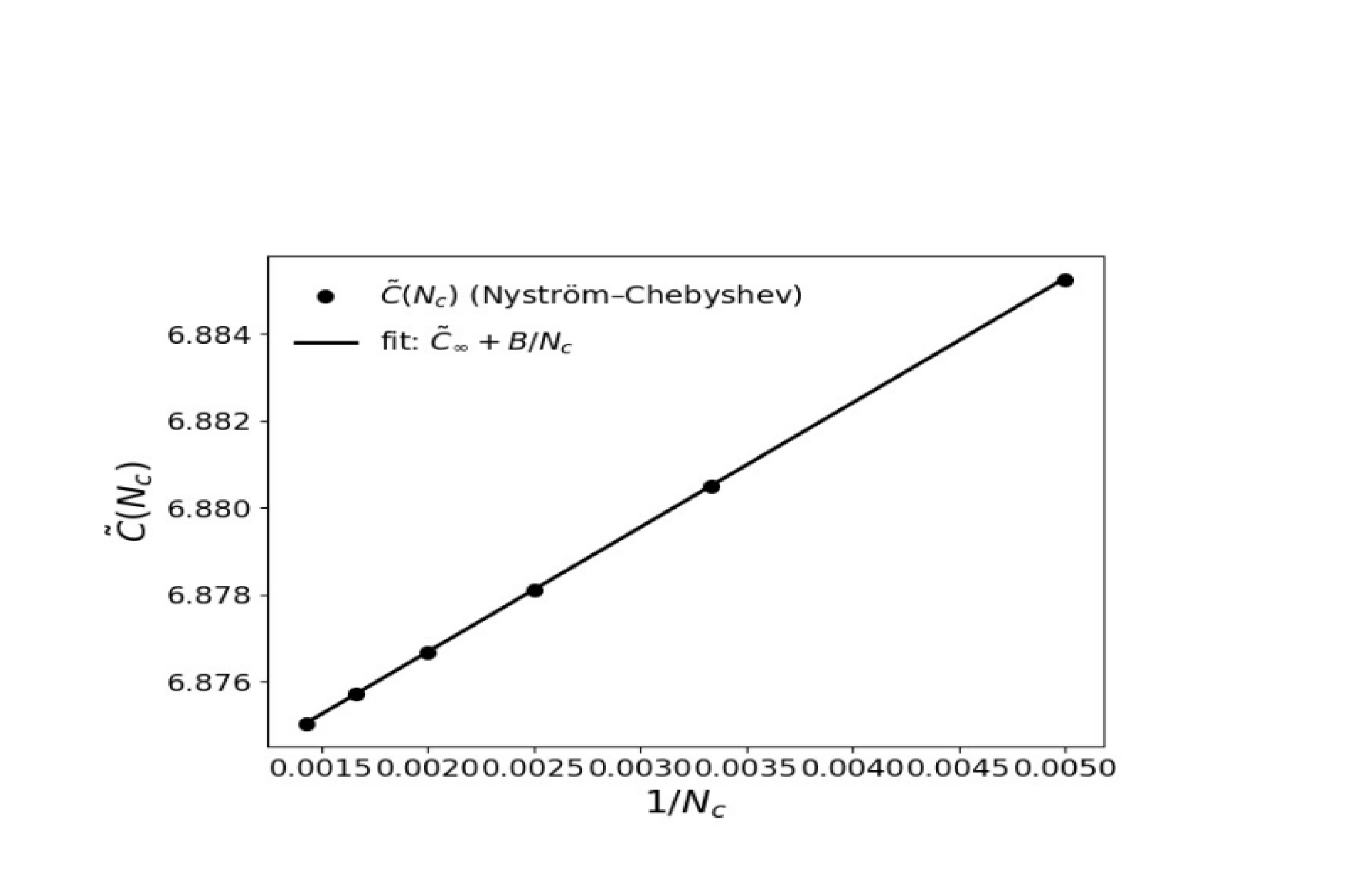}%

\caption{Finite-size scaling of the dimensionless capacitance $\tilde C(N_c)$ computed using $N_c$ Gauss--Chebyshev nodes per surface in the Nystr\"om--Chebyshev discretization. The linear dependence on $1/N_c$ supports the asymptotic form $\tilde C(N_c)=\tilde C_{\infty}%
+B/N_c$, yielding the extrapolated continuum estimate $\tilde C_{\infty}%
=6.870944116657182$.}
\end{figure}

\end{document}